\documentclass[preprint,showpacs]{revtex4}
\usepackage{epsfig}
\usepackage[nooneline]{subfigure}
\usepackage{amsmath}
\usepackage{caption}

\begin{document}

\author{D. Froemberg}
\affiliation{Institut f\"ur Physik,
Humboldt-Universit\"at zu Berlin, Newtonstra\ss e 15, 12489 Berlin, Germany}
\author{H. H. Schmidt-Martens}
\affiliation{Institut f\"ur Physik,
Humboldt-Universit\"at zu Berlin, Newtonstra\ss e 15, 12489 Berlin, Germany}
\author{I.M. Sokolov}
\affiliation{Institut f\"ur Physik,
Humboldt-Universit\"at zu Berlin, Newtonstra\ss e 15, 12489 Berlin, Germany}
\email{e-mail: igor.sokolov@physik.hu-berlin.de}
\author{F. Sagu\'es}
\affiliation{Departament de Quimica Fisica, Universitat de Barcelona, Mart{\'\i} i
Franqu\'es 1, E-08028 Barcelona, Spain}

\title{Asymptotic front behavior in an $A+B\rightarrow 2A$ reaction under subdiffusion}

\begin{abstract}
We discuss the front propagation in the $A+B\rightarrow 2A$ reaction under
subdiffusion which
is described by continuous time random walks with a heavy-tailed power law waiting
time probability
density function. Using a crossover argument, we discuss the two scaling regimes of the
front propagation: an intermediate asymptotic regime given by the front solution of
the 
corresponding continuous equation, and the final asymptotics, which is
fluctuation-dominated and therefore lays out of reach of the continuous scheme. 
We moreover show that the continuous reaction subdiffusion equation indeed possesses a 
front solution that decelerates and becomes narrow 
in the course of time. This continuous description breaks down for larger times when
the front
gets atomically sharp. We show that the velocity of such fronts 
decays in time faster than in the continuous regime.
\end{abstract}

\maketitle

\section{Introduction}

Reactions under subdiffusion have attracted much attention in recent years due to
their growing relevance
for description of processes taking place in porous media such as certain geological
formations or gels,
in the crowded cell interiors and in many other strongly inhomogeneous environments
including 
modern drug delivery systems. We focus here on the autocatalytic conversion
$A+B\rightarrow2A$, a reaction that exhibits travelling 
front solutions if the initial conditions are chosen appropriately, i.e. if $A$ and
$B$ are initially 
separated in space \cite{KPP, Murray}.

We concentrate on situations when subdiffusion can be modelled within the CTRW
scheme with a waiting time 
probability density function (pdf) decaying
according to a power law, $\psi(t)\propto t^{-1-\alpha}$. The continuous description
of the $A+B\rightarrow2A$ 
reaction under subdiffusion, following locally the mass action law corresponding to
the FKPP equation, was derived in
\cite{trfront1} and is given by a partial integro-differential equation with a
kernel depending on the particle 
concentrations at all times.

In that preceding work we have shown analytically that the resultant minimal front
velocity
goes to zero under the assumption of a constant front shape, which was interpreted
as propagation failure.
In a following paper \cite{trfront2}, numerical simulations corroborated this
picture, while two different regimes of
front propagation were identified. In the fluctuation dominated regime, pertinent to
large reaction rates,
the front velocity was found to decay
as $v(t)\propto t^{\alpha-1}$, whereas in the regime of small reaction rates, for
which the continuous description
applies, the front velocity was observed to go as $v(t)\propto
t^{\frac{\alpha-1}{2}}$. 
Longer simulation runs of the continuous case (small reaction rates) revealed that 
after
an intermediate regime that ranged over less than two orders of magnitude in time where
$v(t)\propto t^{\frac{\alpha-1}{2}}$ applies, the exponent sets in to decay
\cite{CamposMendez09}.
Hence the alleged exponent conjectured from the continuous picture was not the final
one.
Up to now, there has not been any physically sound interpretation of the
front velocities found in these simulations.

In this work we attempt to fill this gap by giving a crossover argument that is used to
construct an Ansatz for the solution of the reaction subdiffusion equation at the
leading edge.
We found that in order to maintain a front velocity
that goes as $v(t)\propto t^{\frac{\alpha-1}{2}}$, the additional assumption 
of the width of the front going as $t^{\frac{\alpha-1}{2}}$ has to be made,
so that the front does not maintain a constant form in the course of its propagation.
Since the front's width decreases with time, any real (or simulated) 
subdiffusive FKPP system will sooner or later undergo a change of regime:
the front will get atomically narrow and the continuous scheme breaks down.
Physically this has to do with the fact that at large times the jump rate
always becomes small compared to the reaction rate, so that the fluctuation
dominated regime sets in.
Since the particles react before they are able to leave the site, the front becomes
atomically sharp.
We suggest that the findings in \cite{CamposMendez09} 
(decay of the exponent characterizing the time dependence of the front velocity) can
be interpreted in the sense of a
transition from the intermediate asymptotics of the reaction described by the
continuous reaction-subdiffusion scheme to the final asymptotics corresponding to
the fluctuation dominated regime.
We start by presenting simple physical arguments in favor of this picture. We then
show that
the intermediate asymptotics with  $v(t)\propto t^{\frac{\alpha-1}{2}}$ indeed
appears as a possible
solution of the corresponding integro-differential reaction-subdiffusion equation.
Physical arguments
show however that this asymptotics cannot be the final one, and that the final
regime is fluctuation-dominated.
We then turn to a numerical investigation of this fluctuation-dominated regime and show
that the subdiffusive nature of the motion leads to additional fluctuation effects
absent in
the normal diffusive case. 

\section{Crossover arguments}

Under normal diffusion and with the overall particle concentration $A+B=c$ being
locally conserved,
the $A+B\rightarrow2A$ reaction is described by the Fisher-Kolmogorov-Petrovskii-
Piscounov (FKPP) 
equation
\[
\frac{\partial A(x,t)}{\partial t} = D \Delta A(x,t) + k(c-A)A
\]
that has been extensively studied in the past. 
According to its classical solution \cite{KPP, Murray}, fronts propagating with
velocities $v \geq \sqrt{2 k cD}$ are possible, and it is moreover known that for 
step-like initial condition the solution with minimum speed, $v=\sqrt{2 k cD}$, is
the one which is really achieved at long times.

In order to gain intuition about the front behavior under subdiffusion, we make use
of the following idea:
for any waiting time pdf $\psi$ with finite mean $\left\langle t\right\rangle $, the
behavior
at very long times $t \gg \left\langle t\right\rangle $ corresponds to normal
diffusion, so that
the behavior pertinent to reaction-diffusion schemes is recovered only if time $t$
is large enough.
On the other hand, if the initial domain of the pdf can be approximated by a power-law,
$\psi(t) \propto t^{-1-\alpha}$ up to some truncation time $T$,
the behavior at short times should correspond to the one in subdiffusion, and there
must be
a smooth crossover from one regime to the other. 
We therefore consider the truncated power-law waiting time distribution
with truncation parameter $T$,
\begin{eqnarray}
\psi_T(t) &=& \frac{ (t_0+T)^\alpha}{(t_0+T)^\alpha-t_0^\alpha} \frac{\alpha
t_0^\alpha }{(t_0+t)^{1+\alpha}} 
\Theta(T-t),
\label{truncPL}
\end{eqnarray}
with mean value
\begin{eqnarray}
\left\langle t\right\rangle &=& \frac{\alpha T t_0^\alpha +
t_0\left(t_0^\alpha-(T+t_0)^\alpha \right) }{(\alpha
-1)\left(t_0^\alpha-(T+t_0)^\alpha \right) }.
\label{mean}
\end{eqnarray}
For $T\gg t_0$, $\left\langle t\right\rangle \approx
\frac{\alpha}{1-\alpha}t_0^\alpha T^{1-\alpha}$.

For small times $t\ll T$, when the system does not feel the cutoff, 
the behavior of the velocity will be similar to that in subdiffusion,
whereas for large times the behavior will be the classical one with a constant
minimal velocity.
The crossover between the two regimes must thus take place at some crossover time
$t_{cr}$.
We assume that in the anomalous domain $v \propto t^\beta$,
and that after this a crossover to normal behavior sets in. 
In the case when the normal behavior is described by the FKPP scheme this
corresponds to $v = const. \sim \sqrt{c kD}$,
with $D = a^2/2\langle t \rangle$, where $a$ is the step's length of the corresponding
random walk process (an irrelevant microscopical variable), and the time
behavior of the velocity in the anomalous regime is given by the equation
\begin{equation}
t_{cr}^\beta \simeq \left[ ck \frac{a^2}{2 \langle t(t_{cr}) \rangle}\right]^{1/2}
\label{eq:vel}
\end{equation}

In order to determine the crossover time we concentrate on the most basic quantity
that is known in the normal as well as in the anomalous case, i.e.
the number of performed steps, a measure of mobility, which is given by
\begin{eqnarray}
n_{D}(t) &=& \frac{t}{\left\langle t\right\rangle} 
\end{eqnarray}
in the normal regime $t\gg t_{cr}$, and
\begin{eqnarray}
n_{SD}(t) &=& \frac{t^\alpha}{\Gamma[1+\alpha]t_0^\alpha} 
\end{eqnarray}
in the subdiffusive regime $t\ll t_{cr}$.

By enforcing $n_{SD}(t_{cr}) = n_{D}(t_{cr})$ we find
\begin{equation}
\frac{1-\alpha}{\alpha}\frac{t_{cr}}{t_0^\alpha T^{1-\alpha}} =
\frac{t_{cr}^\alpha}{\Gamma(1+\alpha)t_0^\alpha}; 
\end{equation}
and hence $t_{cr}\propto T$
(more precisely
$t_{cr}^{1-\alpha}=\frac{\alpha}{\Gamma(1+\alpha)(1-\alpha)}T^{1-\alpha}$).
Obviously, the larger we choose the cutoff-parameter $T$, the larger becomes the
crossover time.
At the time the crossover takes place, the quantities characterizing the behavior of
the system,
such as the number of performed steps, the front velocities etc. have to match for
the two
regimes. Tuning $T$ we get the respective values of the quantities of interest at
$t_{cr}$, for example
the mean waiting time $\left\langle t\right\rangle \propto t_{cr}^{1-\alpha}$
for the normal case in terms of $t_{cr}$. From Eq.(\ref{eq:vel}) we then get
\begin{eqnarray}
v(t < t_{cr}) \propto t^{\frac{\alpha -1}{2}}
\label{Vclass}
\end{eqnarray}
in the subdiffusive regime. Correspondingly we can define other time-dependent
effective characteristics in the anomalous regime, e.g. an
effective mean waiting time,
$\left\langle t\right\rangle _{ef f} \propto t^{1-\alpha}$ (the parameter $t_{cr}$
is changed to $t$) which yields
an effective, 
time dependent diffusion coefficient $D_{ef f} \propto 1/t^{1-\alpha}$, from which
Eq.(\ref{Vclass}) can be obtained via the classical formula $v=\sqrt{2ckD_{eff}}$.
This discussion elucidates the source of the anomalous
front velocity in the regime
of small reaction rates, as found numerically in \cite{trfront2}.

We note that even the case for normal diffusion is not simple at all, especially when
the one-dimensional situation is considered, the one especially prone to fluctuation
effects.
To understand the situation we first recall that the FKPP equation, if it holds, has
the same form
in whatever spatial dimension, and provides us not only with the velocity of the
front, but also with the front's
width. Since in any spatial dimension $d$ the dimensions of the concentration $[c]=
\mathrm{L}^{-d}$
and that of the reaction rate $[k] = T^{-1}[c]^{-1}$ are connected to each other, so
that 
$[kc]$ always has the dimension of the inverse time, the combination $\sqrt{Dkc}$
always has
the dimension of velocity, and the combination $w=\sqrt{D/kc}$ always has the
dimension of length.  
The characteristic width of the front 
is thus proportional to our parameter $w$, see \cite{Murray} for a quantitative
discussion.
The velocity of the front and its width $w$ are connected by a simple relation
\begin{equation}
v \sim w/\tau = D/w 
\label{VG}
\end{equation}
where $\tau=w^2/D$ is of the order of the time which it takes a particle to diffuse
through the
front's width. 

Here it is important to note, that the width
$w$ is the only relevant parameter of the dimension of length in the continuous
theory, but
going to the particle picture, another characteristic length, the interparticle
distance
$l = c^{-1/d}$ emerges, and an additional dimensionless parameter $\Pi = w/l$ appears.
The parameter $\Pi$ gives us the front width measured in the units of the
interparticle distance, and
quantifies the strengths of fluctuation effects in the $A+B \to 2B$ reaction. 

According to the Buckingham's $\Pi$-theorem, 
the velocity $v$, $[v] = \mathrm{L} \mathrm{T}^{-1}$ has to depend on the parameters
of the problem as
\[
v = \sqrt{k cD} f(\Pi) = \sqrt{k cD}f\left(\sqrt{D/k} c^{1/d-1/2} \right),
\]
with $f$ being a yet unknown function of a dimensionless parameter, and the
prefactor of $f$
reproducing the classical FKPP behavior of the velocity. The prefactor of $f$ has
the same form in
any spatial dimension, while 
the dimensionless argument of the function $f$ has different form in spaces of
different dimension.
Evidently, the continuous description only works if
$w \gg l$, i.e., in the classical case where $\sqrt{D /c k} \gg c^{-1/d}$: for large
concentrations and diffusion
coefficients and for small reaction rates. In this case there are many particles within
the front region, and the continuous description does hold. For $\Pi \sim 1$
corresponding to the
atomically sharp front, the number of particles across the region fluctuates strongly, 
and therefore front propagation is fluctuation dominated.

Let us now concentrate on the one-dimensional case, as discussed in
\cite{trfront2} and \cite{CamposMendez09}. The fluctuation dominated regime in 1d
corresponds 
to $v \propto Dc$ \cite{Mai96}, which can be easily understood within Eq.(\ref{VG})
by assuming 
the width of atomically sharp front to correspond to the interparticle distance, $w
\simeq l=c^{-1}$.
Repeating the same crossover arguments, as in the previous case, this kind
of behavior under normal diffusion is mirrored onto the form 
\begin{equation}
v(t)\propto t^{\alpha-1}
\label{Vfd}
\end{equation} 
for the velocity time dependence in the subdiffusive case.

The same crossover arguments as applied to the velocity, can be also extended to the
width of the
front. Since the front width $w \propto D_{eff}^{1/2}$ is a decaying function of
time in the subdiffusive case,
the condition for continuous description to hold breaks down for times long enough,
and the transition from the intermediate ``classical'' asymptotics, Eq.(\ref{Vclass}),
to the final fluctuation-dominated asymptotics Eq.(\ref{Vfd}) inevitably takes place.

In what follows we first show that the ``classical'' asymptotics, Eq.(\ref{Vclass}),
indeed
appears as a possible solution of the reaction subdiffusion equation, and then we
change to
investigating the far asymptotic regime, when the reaction-subdiffusion equation
breaks down. 
This is done by use of extensive numerical simulations.

\section{Continuous reaction-subdiffusion regime}

Let us assume the front to behave in accordance with our crossover arguments, namely
to have 
the velocity and the width going as $t^{\frac{\alpha-1}{2}}$
(i.e. with position $x(t) \propto v_0 t^{\frac{1+\alpha}{2}}$). The overall form of
the front will be assumed 
exponential at its leading edge $x \to \infty$. Thus, the following Ansatz is made:
\begin{equation}
A(x,t) =
A_0\exp\left[-\lambda_0t^{\frac{1-\alpha}{2}}\left(x-v_0t^{\frac{1+\alpha}{2}}\right)\right] =
A_0\exp\left[-\lambda_0t^{\frac{1-\alpha}{2}}z \right],
\label{ansatzFront}
\end{equation}
where $z=x-v_0t^{\frac{1+\alpha}{2}}$ is the comoving variable.
(The exponential Ansatz is due to the fact that we will anyhow linearize the
equations at the front's far edge,
and we know from elsewhere \cite{APPB2}
that the (stationary) solutions of linear reaction-subdiffusion equations are
exponentials.)

The equation for the concentration of A-particles $A(x,t)$, with $c$ being the
overall particle concentration,
is (cf. \cite{trfront1})
\begin{eqnarray}
\frac{\partial A(x,t)}{\partial t} &=&  k(c-A(x,t))A(x,t) + \frac{a^2}{2}\Delta
\int_0^t M(t-t^\prime) \nonumber \\ && \times (A(x,t^\prime)-c) \exp
\left[-\int_{t^\prime}^t k A(x,t^{\prime\prime})dt^{\prime\prime}\right]dt^\prime.
\end{eqnarray}

We note that $A(x,t)$ becomes small at the leading edge $x\to\infty$, and
$\exp\left[ -\int_{t^\prime}^t k A(x,t^{\prime\prime})dt^{\prime\prime}\right]
\approx 1$, so that
\begin{eqnarray}
\frac{\partial A(x,t)}{\partial t} &=& \frac{a^2}{2}\int_0^t \Delta\left\{
M(t-t^{\prime})(A(x,t^{\prime})-c)
\exp\left[-k \int_{t^{\prime}}^t A(x,t^{\prime\prime})\,dt^{\prime\prime}\right]
\right\}dt^{\prime} \nonumber \\ 
&& + k(c-A(x,t))A(x,t) \label{basicfkpp} \\
&\approx&
\frac{a^2}{2}\int_0^t
M(t-t^{\prime})\Bigg[ \Delta A(x,t^{\prime}) - 
2\nabla A(x,t^{\prime})\int_{t^\prime}^t  k\nabla
A(x,t^{\prime\prime})\,dt^{\prime\prime}\nonumber \\
&&
+(c-A(x,t^\prime))\int_{t^\prime}^t k\Delta A(x,t^{\prime\prime}) \,dt^{\prime\prime} 
- (c-A(x,t^{\prime}))\Big(\int_{t^\prime}^t k\nabla
A(x,t^{\prime\prime})\,dt^{\prime\prime}\Big)^2 \Bigg]
dt^{\prime} \nonumber \\ 
&& + k(c-A(x,t))A(x,t) \label{SD-FKPPlin}
\end{eqnarray}
In particular, with Ansatz (\ref{ansatzFront}) and taking into account that the term
$t^{-\frac{1+\alpha}{2}}$ is negligible
for large $t$, we have
\begin{eqnarray}
\frac{\partial A(x,t)}{\partial t} &=& A_0 \exp\left[ -\lambda _0
t^{\frac{1-\alpha}{2}}\left( x-v_0 t^{\frac{1+\alpha}{2}} \right) \right]\times
\nonumber \\
&&\left[ v_0\lambda _0
t^{\frac{1-\alpha}{2}}t^{\frac{\alpha-1}{2}}\frac{\alpha+1}{2} - 
\lambda_0 t^{-\frac{1+\alpha}{2}}\frac{1-\alpha}{2}\left(
x-v_0t^{\frac{1+\alpha}{2}}\right) \right]\nonumber \\
&=& A_0 \exp\left[ -\lambda _0 t^{\frac{1-\alpha}{2}}\left( x-v_0
t^{\frac{1+\alpha}{2}} \right) \right]\times \nonumber \\
&& \left[v_0\lambda_0 -\frac{1-\alpha}{2}\lambda _0 x t^{-\frac{1+\alpha}{2}}
\right] \nonumber \\
&\approx& A_0  \exp\left[-\lambda_0t^{\frac{1-\alpha}{2}} (x-v_0
t^{\frac{\alpha+1}{2}}) \right]
v_0\lambda_0 \label{lhs} \\
\nabla A(x,t) &=& -A_0 \lambda _0 t^{\frac{1-\alpha}{2}} 
\exp\left[ -\lambda _0 t^{\frac{1-\alpha}{2}}\left( x-v_0 t^{\frac{1+\alpha}{2}}
\right) \right] \nonumber \\
\Delta A(x,t) &=& A_0 \lambda _0^2 t^{1-\alpha} \exp\left[ -\lambda _0
t^{\frac{1-\alpha}{2}}\left( x-v_0 t^{\frac{1+\alpha}{2}} \right) \right]. \nonumber
\end{eqnarray}

Proceeding as in \cite{trfront1} we have to first order in concentration for the $\mathrm{A}$-particles:
\begin{eqnarray}
\frac{\partial A(x,t)}{\partial t}  &\approx&
\frac{a^2}{2} \int_0^t M(t-t^\prime) \Delta
A_0\exp\left[-\lambda_0t^{\prime\frac{1-\alpha}{2}} (x-v_0
t^{\prime\frac{\alpha+1}{2}}) \right]  \,dt^\prime  \nonumber \\
&& +\frac{a^2}{2}
\int_0^t M(t-t^\prime) c k 
\int_{t^\prime}^t \Delta A_0
\exp\left[-\lambda_0t^{\prime\prime\frac{1-\alpha}{2}}\left(x-v_0t^{\prime\prime\frac{1+\alpha}{2}}\right)\right]
\,dt^{\prime\prime}
\,dt^\prime \nonumber \\
&& +c k A(x,t), \label{A-linOp}
\end{eqnarray}
i.e.
\begin{eqnarray}
&&A_0  \exp\left[-\lambda_0t^{\frac{1-\alpha}{2}} (x-v_0 t^{\frac{\alpha+1}{2}})
\right]v_0\lambda_0 \nonumber \\
&\approx&
\frac{a^2}{2} \int_0^t M(t-t^\prime)  A_0 \lambda _0^2 t^{\prime 1-\alpha}
\exp\left[-\lambda_0t^{\prime\frac{1-\alpha}{2}} (x-v_0
t^{\prime\frac{\alpha+1}{2}}) \right]
\,dt^\prime  \nonumber \\
&& +\frac{a^2}{2}
\int_0^t M(t-t^\prime) c k A_0 \lambda _0^2 
\int_{t^\prime}^t {t^{\prime\prime}}^{1-\alpha}
\exp\left[-\lambda_0t^{\prime\prime\frac{1-\alpha}{2}}\left(x-v_0t^{\prime\prime\frac{1+\alpha}{2}}\right)\right]
\,dt^{\prime\prime}
\,dt^\prime \nonumber \\
&& +c k A_0\exp\left[-\lambda_0t^{\frac{1-\alpha}{2}} (x-v_0 t^{\frac{\alpha+1}{2}})
\right] , \label{A-lin}
\end{eqnarray}
with the kernel
\begin{equation}
\tilde M(u) = \frac{u\tilde\psi(u)}{1-\tilde\psi(u)}  \nonumber
\end{equation}
in Laplace domain (which corresponds to the Riemann-Liouville fractional derivative
of order $1-\alpha$
in the subdiffusive case, $\frac{1}{\Gamma(\alpha)}\frac{d}{dt}\int_0^t
\frac{1}{(t-t^{\prime})^{1-\alpha}} (\cdot)  dt^\prime$).

We note that in the following we assume $\psi(t)\propto \tau^\alpha t^{-1-\alpha}$ so that
the new parameter $\tau$  and the old one $t_0$ from the original waiting time
distribution
$\psi(t) = \frac{\alpha t_0^{\alpha} }{(t+t_0)^{1+\alpha}}$ (i.e. the $\psi$ we
truncated for the crossover argumentation
in the preceding section cp. (\ref{truncPL})) turn out to be the same, $\tau=t_0$.

Altogether we have then for $z=x-v_0t^{\frac{1+\alpha}{2}}$ and $t$ large:
\begin{eqnarray}
&&\lambda_0 v_0 \exp\left[-\lambda_0t^{\frac{1-\alpha}{2}}z\right] \nonumber \\
= && \exp\left[-\lambda_0t^{\frac{1-\alpha}{2}}z\right]
\left[ \frac{a^2}{2 \Gamma(\alpha )\Gamma (1-\alpha )\tau^\alpha}\left[ B \lambda_0^2 +
\frac{c k\lambda_0}{v_0}\left[1-B\right]  \right] +c k \right],
\end{eqnarray}
where $B$ is a constant that originates from the estimation of the involved
integrals, see
Appendix \ref{app:intCalc}, with $B(\alpha,2-\alpha)\geq B\geq0$ and $B(\nu,\mu)$
being the Beta-function.
This yields the dispersion relation for $\lambda_0$:
\begin{equation}
0 = \lambda_0^2 +\frac{\frac{c k K^*_\alpha}{v_0}\left[ 1-B\right] - v_0}{K^*_\alpha
B} \lambda_0 + \frac{c k}{K^*_\alpha B} \label{disp}
\end{equation}
with $\frac{a^2}{2\Gamma(\alpha)\Gamma(1-\alpha)\tau^\alpha}=K^*_\alpha =
\frac{K_\alpha}{\Gamma(\alpha)}$, where $K_\alpha$ is the generalized diffusion
constant. From
\begin{equation}
\lambda_{0_{1,2}} = -\frac{\frac{c k K^*_\alpha}{v_0}\left[ 1-B\right] -
v_0}{2K^*_\alpha B}
\pm \sqrt{\frac{(\frac{c k K^*_\alpha}{v_0}\left[ 1-B\right] -
v_0)^2}{4K_\alpha^{*2} B^2} - \frac{c k}{K^*_\alpha B}} \label{lambdaeq}
\end{equation}
we find the restriction
\begin{equation}
(\frac{c k K^*_\alpha}{v_0}\left[ 1-B\right] - v_0)^2 \geq 4 c k K^*_\alpha B
,\label{restrictRadi}
\end{equation}
a quartic equation in $v_0$ which yields
\begin{equation}
v_0^2 = K^*_\alpha c k \left[1+B \pm 2\sqrt{B}\right] \label{vquartic}
\end{equation}
Note that in the normal case $B=1$, the minimal front velocity $v_{min} =
\pm2\sqrt{c D k}$
is reproduced; the other solution is a double one at $v=0$ for which there is no front.
Recall again that $B(\alpha,2-\alpha)\geq B\geq0$, therefore eq.(\ref{vquartic})
always has real roots
($B(\alpha,2-\alpha)>1$ for all $\alpha<1$).

This analysis shows that there exists a set of (nonzero) parameters $\lambda_0$ and
$v_0$ for which
Ansatz (\ref{ansatzFront}) yields a solution to the linearized
reaction subdiffusion equation (\ref{SD-FKPPlin}), although the integrals appearing
in  the calculations
can only be estimated approximately. We note that neither an Ansatz taking a front
velocity going as
$v(t)\propto t^{\alpha-1}$ nor an Ansatz with $v(t)\propto t^\frac{\alpha-1}{2}$ and a 
constant front width yield an asymptotic solution of the reaction-subdiffusion
equation, and therefore
such types of behavior are impossible within the continuous scheme. 

In our previous simulations we were not able to detect the changes in the front
shape, presumably due to our averaging
procedure over several runs, and hence did not conjecture any change of regime in
\cite{trfront2}.
On the other hand, our simulations were not carried out for long enough times to
detect the change of regime
in the velocity variable. Since such transitions take place only very slowly, much
longer runs 
of the simulation were really necessary,
as the more extensive simulations of \cite{CamposMendez09} showed. This suggests
that indeed the continuous regime
as considered above does not describe the final behavior of the front. Now we can
interpret the 
findings of Ref. \cite{CamposMendez09} as the setting in of a slow transition
to the fluctuation dominated regime.

\section{Failure of the continuous description: Atomically sharp fronts in
simulations for large times}

Since the subdiffusive front is slowing down and becoming steeper in the course of
time,
any system will sooner or later enter a regime already discussed in \cite{trfront2}
for subdiffusion
and in \cite{Mai96, Mai00} for normal diffusion. This regime is a fluctuation
dominated one and is 
no longer described by continuous approaches. Since the integral kernel $M(t)$ of the 
linearized reaction-subdiffusion equations decays and determines the mean density of steps in time,
the waiting times for particles at a site
become so large in the course of time that the motion of the front is governed by the 
first $\mathrm{A}$-particle entering a new site. All
$\mathrm{B}$-particles at the same site have enough time to react with $\mathrm{A}$
before the next jump from the site takes place, the reaction rate dependence
disappears,
and the behavior of the front gets to be the same as in the reaction on the first
contact.

Under such a condition the velocity of the front's motion can be estimated using the
following argument (adapted from \cite{SanderFront00,SanderFront01} for our sequential updating scheme). 
Let us consider the front position as fixed by the rightmost
A-particle(s), and concentrate on the next jump of the front particle. If the
A-particle is
alone at its front position, this next jump takes place with probability 1/2 by an
amount $\pm a$, so that
the net front displacement after such a step is zero on average. On the contrary, if
there
is more than one particle at the front position (the probability of which is $ac$ if
the concentration
is defined as a number of particles per unit length) the front
moves by $a$ to the right, if the particle makes a step forward (which happens with
probability 1/2), and does not move, if it jumps backwards, since then there is at
least one other 
particle, which keeps the front position where it was. Therefore, at a step of a
front particle, 
the front moves on average by a distance $a^2 c/2$. Since the rate at which the
particle moves is
defined by the time-integral of the memory kernel $M$, the front's velocity is given by
\begin{equation}
v \approx \frac{a^2 c}{2} \int_0^t M(t-t^\prime) dt^\prime.
\label{SharpF}
\end{equation}
Let us first derive the asymptotic jump rate of the particles. 
Consider the generic waiting time pdfs with the asymptotic behavior
\begin{equation}
\psi(t) \propto \tau^\alpha t^{-1-\alpha}.
\end{equation}
The (cumulative) probability to make a step until $t$, for $t$ large is then
\begin{equation}
\Psi(t) \simeq 1-\tau^\alpha t^{-\alpha};
\end{equation}
or in Laplace domain, using the Tauberian theorem
\begin{equation}
\tilde \Psi(u) \simeq \frac{1}{u} - \Gamma(1-\alpha)\tau^\alpha u^{-1+\alpha},
\end{equation}
so that the pdf
\begin{equation}
\tilde \psi (u) \simeq 1 - \Gamma(1-\alpha)\tau^\alpha u^{\alpha}.
\label{thepdf}
\end{equation}
The rate for a particle to jump is $\int_0^t M(t-t^\prime) dt^\prime$ or in Laplace
domain
\begin{equation}
\frac{\tilde M(u)}{u} = \frac{\tilde\psi(u)}{1-\tilde\psi(u)} \simeq
\frac{1}{\tau^\alpha \Gamma(1-\alpha)} u^{-\alpha}
\end{equation}
for $u\to 0$ so that we have an expression for the velocity in the Laplace domain
given by
\begin{equation}
\mathcal{L}\left\lbrace v(t)\right\rbrace  =
\frac{ca^2}{2}\frac{1}{\tau^\alpha \Gamma(1-\alpha)} u^{-\alpha}.
\end{equation}
Transforming back to the time domain yields
\begin{equation}
v(t)  = \frac{a^2}{2 \Gamma(1-\alpha)\tau^\alpha} \frac{c}{\Gamma(\alpha)} t^{\alpha-1}
=  K_\alpha \frac{c}{\Gamma(\alpha)} t^{\alpha-1}= c K_\alpha^*t^{\alpha-1}.
\end{equation}
With
$\frac{1}{\Gamma\left( \alpha\right)\Gamma\left(1-\alpha \right)} =
\frac{\sin\left(\alpha\pi \right) }{\pi}$
the front velocity is better expressed as
\begin{equation}
v(t) = \frac{a^2}{\tau^\alpha}\frac{c}{2} \frac{\sin\left(\alpha\pi \right) }{\pi}
t^{\alpha-1}
\label{v_theor}
\end{equation}
which corresponds to the position of the front going as 
\begin{equation}
x(t)=\frac{N_A}{c} = \int_0^t v(t)dt =
\frac{a^2}{2\tau^\alpha}\frac{\sin\left(\alpha\pi \right)}{\alpha \pi} c t^{\alpha}.
\label{NA_theor}
\end{equation}
($N_A$ is the total amount of A-particles). 

\begin{figure}[h]
\includegraphics[width=.5\textwidth]{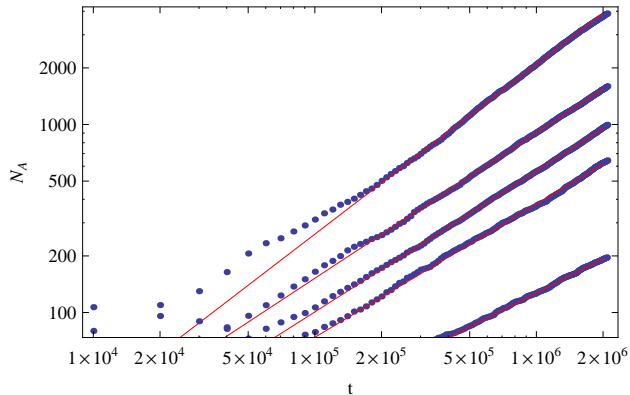}
\caption{\label{fluctfront} Front position for $\alpha = 0.9,\;0.8,\;0.75,\;0.7,\;0.6$
(upper to lower graphs), $c=0.3$.
Red lines denote fits of the large time behavior.}
\end{figure}

Note that the definition of the characteristic waiting time $\tau$ adopted here does not 
allow for simply taking $\alpha=1$ to perform the limiting transition to normal diffusion,
as found e.g. for the exponential distribution of waiting times, 
$\psi(t) = \langle t \rangle^{-1}\exp(-t/\langle t \rangle)$. This is due to the presence of the divergent 
$\Gamma(1-\alpha)$ in Eq.(\ref{thepdf}): the case $\alpha=1$ corresponds, strictly speaking to 
still (logarithmically) divergent mean waiting times. For the normal case with
converging mean Eq.(\ref{thepdf}) reads $\tilde \psi (u) \simeq 1 - \langle t \rangle u$,
and, after performing the same steps as above, the front velocity of the normal
fluctuation dominated regime,
$v_{fluct}= cD$, with $D$ being the diffusion constant, is recovered \cite{Mai96}.
Fig. \ref{fluctfront} shows the total number of particles in the simulation for the
fluctuation
dominated regime, i.e. reaction on contact, for a concentration $c=0.3$.
In these simulations we had
$7$ runs for $\alpha=0.9$, $18$ for $\alpha=0.8$, $41$ for $\alpha=0.75$,
$13$ for $\alpha=0.7$ and $18$ for $\alpha=0.6$.

Table \ref{tabexp} shows the exponents of the long time fits $N_A = F t^\beta$ which
coincide well with
$\alpha$.

\begin{table}[hb]
\centering
\begin{tabular}{|l||c|c|c|c|c|}
\hline
$\alpha$ & $0.6$ & $0.7$ & $0.75$ & $0.8$ & $0.9$ \\
\hline 
$\beta$ & $0.603\pm0.004$ & $0.708\pm0.004$ & $0.750\pm0.001$ & $0.775\pm0.002$  &
$0.890\pm0.009$ \\
\hline
\end{tabular}
\caption{ Exponents for the fit $N_A =Ft^{\beta}$ for different $\alpha$. }
\label{tabexp}
\end{table}

The values of the prefactor $F$ found from the simulations turned out to be however
larger than the predicted ones in
(\ref{NA_theor}) by around $30-40 \%$. In order to find out about the origin of this
difference, we 
performed simultaneous simulations of subdiffusion and of subdiffusion with
randomized particles, i.e.
in the situation when the particles lost their individual memory and were chosen
randomly to jump when a jumping time was reached.
This variant of the reaction closely mimics the behavior assumed to derive
Eq.(\ref{SharpF}),
namely the assumption that the rate at which the steps of the rightmost A particle
are made is equal to the mean jump rate of all particles at time $t$: we fully
disregard the fact that the rightmost A is a very special particle, with its special
prehistory. 

\begin{figure}[h]
\includegraphics[width=.5\textwidth]{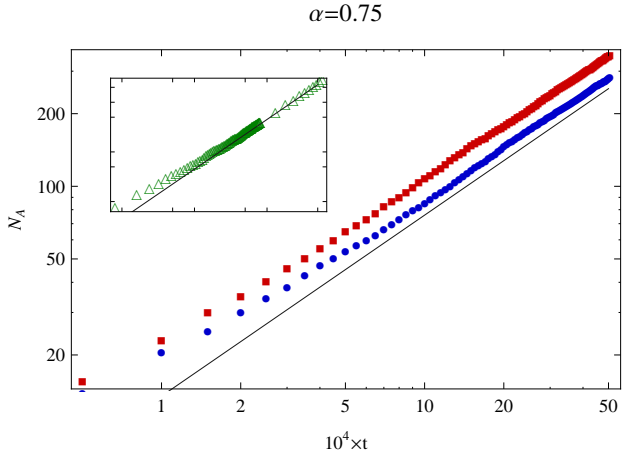}
\caption{\label{front-t} Time dependence of the total amount of A-particles $N_A$ for 
the subdiffusive case (squares) and subdiffusion with
randomized particles (circles). The black line denotes the theoretical curve
according to (\ref{NA_theor}). The inset shows the situation for
an exponential waiting time pdf (with mean $1$), $t$ goes from $10$ to $5\times
10^{4}$, $N_A$ goes from $6$ to $2000$. The black line denotes again the theory,
$N_A = D c^2 t$; $c=0.3$.}
\end{figure}

Fig. \ref{front-t} shows the time dependence of the overall amount of A-particles
for $\alpha=0.75$.
The theoretical curve (\ref{NA_theor}) lies much closer to the simulation results of
subdiffusion with
randomized particles. The remaining difference between the simulation of the
randomized particles
and the theoretical result is presumably due to the fact that convergence to the
asymptotic behavior  in subdiffusion
is very slow. Apparently, the full subdiffusive picture implies an additional
fluctuation effect.
For a better interpretation of the results, we also simulated
the case of normal diffusion. The inset of the figure shows the situation for
an exponential waiting time pdf with mean $1$,
where the simulated front behavior converges to the predicted behavior indicated by the
black line, $N_A = D c^2 t$. We note that Warren et al. \cite{SanderFront00}
detected a fluctuation
effect in the normal case that occurs at small concentrations. However, as the inset
shows,
due to the sequential update in our simulations, this effect does not come into play
here and our
theoretical approach is sufficient to explain the front behavior in the normal case.

\begin{figure}[h]
{\includegraphics[width=.5\textwidth]{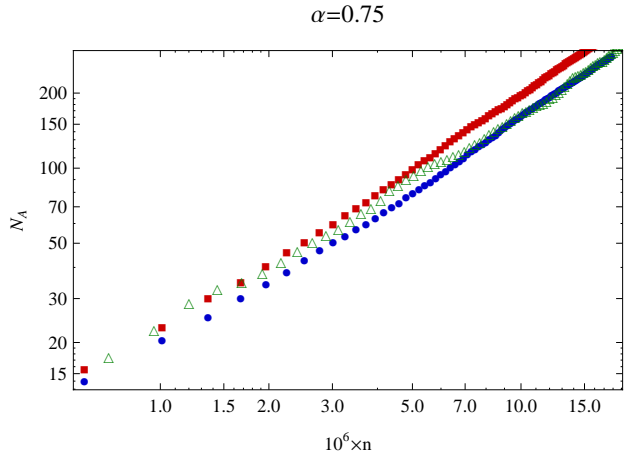}}\\
\caption{\label{front-n} Total amount of A-particles $N_A$ for the normal case
(triangles),
the subdiffusive case (squares) and subdiffusion with
randomized jumps (circles), both $\alpha=0.75$, depending on
the total number of performed steps; $c=0.3$.}
\end{figure}

\begin{figure}[h]
{\includegraphics[width=.5\textwidth]{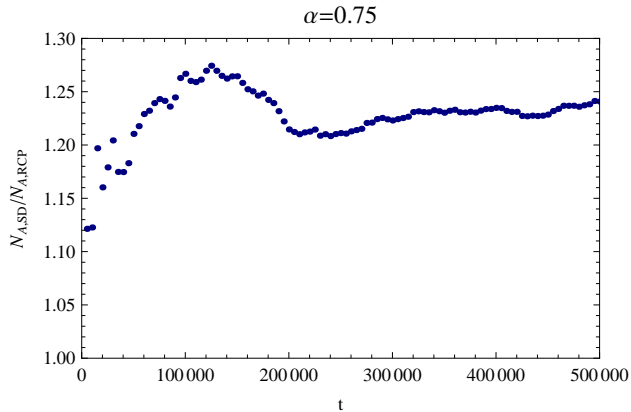}}\\
\caption{\label{quot-t} Quotient of total amount of A-particles for 
the subdiffusive case and subdiffusion with randomized particles
$\frac{N_{A,SD}}{N_{A,RCP}}$
as a function of time. }
\end{figure}

Fig. \ref{front-n} shows the dependence of the overall amount of A-particles on
the total amount of steps performed for $\alpha=0.75$.
Comparing the two subdiffusive prescriptions (original and randomized)
as well as the normal diffusion reveals that the randomized version of subdiffusive
front behavior
is more akin to the normal diffusive front behavior than the full subdiffusive version:
If we interpret the number of steps $n$ as the internal, operational time of the
process, the randomized
subdiffusive setting and the normal diffusive one have the same asymptotics, whereas
the full original subdiffusive front position differs by a certain factor.
Fig. \ref{quot-t} shows the quotient of the original subdiffusive front position and
the randomized one,
which can be used to quantify this effect that turns out to be around at least
$20-30 \%$.

Obviously, the additional fluctuation effect of the front behavior is genuinely due
to subdiffusion.
This effect cannot be explained within the mean-field description of the front
behavior,
but comes into play through the interaction of the particles at the front: The rate
at which a front
particle performs a jump is higher than the average jump rate of a single particle
in the system.
If the particle at the edge of the front is subject to a very long waiting time
(which happens not often, but occasionally), other particles will outpace that
particle and take the lead.
Hence, the impact of very long waiting times in single particle dynamics on the
front motion is 
considerably reduced.

\section{Conclusions}

We discussed the front motion in the $A+B\rightarrow2A$ reaction under subdiffusion
described by continuous time
random walks where the reaction is governed by the mass action law on a microscopic
scale. 
We have shown that at intermediate times, as long as the process can be described
within a continuous picture,
the front velocity goes as $v(t)\propto t^{\frac{\alpha-1}{2}}$. The decay of the
front velocity goes along
with a decay of the width of the front, which at longer times therefore gets
atomically sharp.
At such times the continuous picture, implied by the description within the
reaction-subdiffusion equations
scheme, inevitably breaks down. The typical time scale of diffusion becomes
very large compared to the typical time scale of reaction, and a crossover to the
fluctuation dominated regime
takes place where the front velocity decays faster, $v(t)\propto t^{\alpha-1}$. This
fluctuation dominated
regime is the same as in the reaction on the first contact, and is characterized by
additional
fluctuation effects compared to the case of normal diffusion.

\appendix

\section{Evaluation of Integrals}\label{app:intCalc}

We investigate the integrals in expression (\ref{A-lin}) term by term, from left to
right and take into account that the constant $A_0$ cancels.
\begin{eqnarray}
I_1 = && \int_0^t M(t-t^\prime)  \lambda _0^2 t^{\prime 1-\alpha}
\exp\left[-\lambda_0t^{\prime\frac{1-\alpha}{2}} (x-v_0
t^{\prime\frac{\alpha+1}{2}}) \right]
\,dt^\prime \nonumber \\
= && \frac{\lambda_0^2}{\Gamma(1-\alpha)\tau^\alpha}
\exp\left[-\lambda_0t^{\frac{1-\alpha}{2}}\left(x-v_0t^{\frac{1+\alpha}{2}}\right)\right]
\times \nonumber \\
&& \frac{1}{\Gamma(\alpha)}\frac{d}{dt}\int_0^t \frac{1}{(t-t^{\prime})^{1-\alpha}}
t^{\prime 1-\alpha}
\exp\left[-\lambda_0t^{\prime\frac{1-\alpha}{2}}\left(x-v_0t^{\prime\frac{1+\alpha}{2}}
\right) + 
\lambda_0t^{\frac{1-\alpha}{2}}\left(x-v_0t^{\frac{1+\alpha}{2}} \right)\right]
 dt^\prime.\nonumber \\ 
\label{int1}
\end{eqnarray}
This expression can be estimated from above since $t^\prime\leq t$:
\begin{eqnarray}
I_1 \leq && \frac{\lambda_0^2}{\Gamma(1-\alpha)\tau^\alpha}
\exp\left[-\lambda_0t^{\frac{1-\alpha}{2}}\left(x-v_0t^{\frac{1+\alpha}{2}}\right)\right]
\frac{1}{\Gamma(\alpha)}\frac{d}{dt}\int_0^t \frac{1}{(t-t^{\prime})^{1-\alpha}}
t^{\prime 1-\alpha} dt^\prime \nonumber \\
= && \frac{\lambda_0^2}{\Gamma(1-\alpha)\tau^\alpha}
\exp\left[-\lambda_0t^{\frac{1-\alpha}{2}}\left(x-v_0t^{\frac{1+\alpha}{2}}\right)\right]
\frac{1}{\Gamma(\alpha)}\frac{d}{dt} t \int_0^1 \frac{1}{(1-t^{\prime})^{1-\alpha}}
t^{\prime 1-\alpha} dt^\prime \nonumber \\
= && \frac{\lambda_0^2}{\Gamma(1-\alpha)\tau^\alpha}
\exp\left[-\lambda_0t^{\frac{1-\alpha}{2}}\left(x-v_0t^{\frac{1+\alpha}{2}}\right)\right]
\frac{1}{\Gamma(\alpha)}B\left(\alpha, 2-\alpha \right) \label{est_int1}
\end{eqnarray}
the integral in (\ref{int1}) is monotonic, i.e. it must tend to a constant value
$B\leq B\left(\alpha, 2-\alpha \right)$ for large times ($B=1$ for the normal
diffusive case, in particular).\\
We used here the definition of the Beta-function
$B(\mu,\nu)=\frac{\Gamma(\mu)\Gamma(\nu)}{\Gamma(\mu+\nu)}$.

\begin{eqnarray}
I_2 = && \int_0^t M(t-t^\prime) c k \lambda _0^2 
\int_{t^\prime}^t {t^{\prime\prime}}^{1-\alpha}
\exp\left[-\lambda_0t^{\prime\prime\frac{1-\alpha}{2}}\left(x-v_0t^{\prime\prime\frac{1+\alpha}{2}}\right)\right]
\,dt^{\prime\prime}
\,dt^\prime  \label{int2}
\end{eqnarray}

At the far edge of the front,  our comoving variable $z=x-v_0t^{\frac{1+\alpha}{2}}$
is very large. The transition to large $z$ can be achieved by introducing a large
parameter $\gamma$, so that the integral appearing in the integrand of
(\ref{int2}) obtains the form of a Laplace integral which allows for an
asymptotic estimation for $\gamma\to\infty$:
\begin{eqnarray}
&&\begin{array}{c} \\ \lim \\ \scriptstyle{\gamma\to\infty} \end{array}
\frac{\lambda_0^2}{\Gamma(1-\alpha)\tau^\alpha} \int_0^t t^{\prime\prime 1-\alpha}
\exp\left[-\lambda_0t^{\prime\prime\frac{1-\alpha}{2}}\gamma\left(x-v_0t^{\prime\prime\frac{1+\alpha}{2}}
\right)\right]  \,dt^{\prime\prime} \nonumber \\
&=&
\frac{\lambda_0}{v_0 \Gamma(1-\alpha)\tau^\alpha}t^{1-\alpha}
\exp\left[-\lambda_0t^{\frac{1-\alpha}{2}}\left(x-v_0t^{\frac{1+\alpha}{2}}\right)\right],
\end{eqnarray}
that means that for large $\gamma$ the value of the above integral is asymptotically
determined by the
points where the exponent in the integrand attains its maximum, see e.g.
\cite{BleistHand}.\\
Hence, (\ref{int2}) becomes
\begin{eqnarray}
&& \frac{\lambda_0}{v_0}
\exp\left[-\lambda_0t^{\frac{1-\alpha}{2}}\left(x-v_0t^{\frac{1+\alpha}{2}}\right)\right]
\times \nonumber \\
&& 
\Bigg[ 
t^{1-\alpha} \int_0^t M(t-t^\prime) \,dt^\prime - \nonumber \\
&&\int_0^t M(t-t^\prime)t^{\prime 1-\alpha}
\exp\left[-\lambda_0t^{\prime\frac{1-\alpha}{2}}\left(x-v_0t^{\prime\frac{1+\alpha}{2}}
\right) + 
\lambda_0t^{\frac{1-\alpha}{2}}\left(x-v_0t^{\frac{1+\alpha}{2}} \right)\right]
\,dt^\prime
\Bigg] 
\nonumber \\
= && \frac{\lambda_0}{v_0 \Gamma(1-\alpha)\tau^\alpha}
\exp\left[-\lambda_0t^{\frac{1-\alpha}{2}}\left(x-v_0t^{\frac{1+\alpha}{2}}\right)\right]
\times \nonumber \\
&& \left[ \right.
t^{1-\alpha}\frac{1}{\Gamma(\alpha)}\frac{d}{dt}\int_0^t
\frac{1}{(t-t^\prime)^{1-\alpha}}\,dt^\prime - \nonumber \\
&& \frac{1}{\Gamma(\alpha)}\frac{d}{dt}\int_0^t
\frac{1}{(t-t^\prime)^{1-\alpha}}t^{\prime 1-\alpha}  \,dt^\prime
\exp\left[-\lambda_0t^{\prime\frac{1-\alpha}{2}}\left(x-v_0t^{\prime\frac{1+\alpha}{2}}
\right) + 
\lambda_0t^{\frac{1-\alpha}{2}}\left(x-v_0t^{\frac{1+\alpha}{2}} \right)\right]
\left.\right] \nonumber \\
 = &&\frac{\lambda_0}{v_0 \Gamma(1-\alpha)\tau^\alpha}
\exp\left[-\lambda_0t^{\frac{1-\alpha}{2}}\left(x-v_0t^{\frac{1+\alpha}{2}}\right)\right]
\times\frac{1}{\Gamma(\alpha)} \left[ 1 - B \right], \label{est_int2}
\end{eqnarray}
with $B \leq B(\alpha,2-\alpha)$, cf. (\ref{est_int1}).\\

\end{document}